\documentclass[twoside]{article}
\usepackage{fleqn,espcrc2}

\usepackage{graphicx}

\newcommand{\AmS}{{\protect\the\textfont2
  A\kern-.1667em\lower.5ex\hbox{M}\kern-.125emS}}

\title{{\bf Cooper pair dispersion relation in two dimensions}}

\author{S.K. Adhikari,\address{Instituto de F\'{\i}sica Te\'{o}rica, Universidade Estadual 
Paulista, 01405-900 S\~{a}o Paulo, SP, Brazil}
M. Casas,$^{\mbox{\scriptsize b}}$ A. Puente,$^{\mbox{\scriptsize b}}$ A. Rigo,\address{Departament de F\'{\i}sica, Universitat de les Illes Balears, 07071 Palma de Mallorca, Spain}
M. Fortes,$^{\mbox{\scriptsize c}}$ M.A. Sol\'{\i}s,\address{Instituto de F\'{\i}sica, Universidad Nacional Aut\'{o}noma de 
M\'{e}xico, \\ Apdo. Postal 20-364, 01000 M\'{e}xico, DF, M\'{e}xico}
M. de Llano,$^{\mbox{\scriptsize d}}$ A.A. Valladares,\address{Instituto de Investigaciones en Materiales, Universidad Nacional
Aut\'{o}noma de M\'{e}xico, \\ Apdo. Postal 70-360, 04510 M\'{e}xico, DF,
M\'{e}xico}
and O. Rojo\address{PESTIC, Secretar\'{\i}a Acad\'{e}mica \& CINVESTAV, IPN, 04430 M\'{e}xico DF,
M\'{e}xico}
}
       
\begin{document}

\begin{abstract}
 The Cooper pair binding energy {\it vs.} center-of-mass-momentum dispersion relation for
Bose-Einstein condensation studies of superconductivity is found in two
dimensions for a renormalized attractive delta interaction. It crosses over
smoothly from a linear to a quadratic form as coupling varies from weak to
strong.\\

\end{abstract}

\maketitle

For the attractive interfermion potential 
\begin{equation}
V(r)=-v_{0}\delta ({\bf r}),  \label{1}
\end{equation}
where $v_{0}\geq 0$ is the interaction strength, one can apply the
Lippmann-Schwinger as well as the Cooper-pair (CP) equation in two
dimensions (2D) for two fermions of mass $m$ with momenta wavevectors ${\bf k}_{1}$
and ${\bf k}_{2}$ in {\it free space} (i.e., vacuum) and in the momentum space above the 
{\it filled Fermi sea}, respectively. Combining these two equations so as to
eliminate (the regularized, infinitesimally small \cite{GT}) $v_{0}$, one obtains
the {\it renormalized} CP {\it equation} 
\begin{eqnarray}
 \sum_{k}\frac{1}{B_{2}+ {\hbar ^{2}k^{2}}/{m}}-  \qquad \qquad \qquad \quad \qquad \qquad & \nonumber&  \\
 \sum_{k}{}^{\prime }\frac{1}{\hbar
^{2}k^{2}/m+\Delta _{K}-2E_{F}+\hbar ^{2}K^{2}/4m} = 0,  & \label{8} &
\end{eqnarray}
where ${\bf k}\equiv ({\bf k}_{1}-{\bf k}_{2})/2$ is the relative, ${\bf K}%
\equiv {\bf k}_{1}+{\bf k}_{2}$ the center-of-mass momentum (CMM), usually taken as zero in BCS theory, $E_{F}$
the Fermi energy, $B_{2}\geq 0$ the (single-bound-state) pair binding energy
in vacuum, and $2E_{F}-\Delta _{K}$ the total energy of the CP with $%
\Delta _{K}\geq 0$ its binding energy now simply a function of $B_{2}$. \
The prime on the second summation denotes the restriction $|%
{\bf k}\pm {\bf K}/2|>k_{F}$. In 2D one finds that $\Delta_K = B_2$ exactly, but {\it only} for $K \equiv 0$. Expanding Eq. (\ref{8}) for small but nonzero 
$K$ and subtracting from the expression for $K=0$ gives a small-CMM
expansion valid for any dimensionless coupling $B_{2}/E_{F}={\Delta _{0}/E}_{F}$, namely 
\begin{eqnarray}
\varepsilon _{K}\equiv ({\Delta _{0}-\Delta _{K})}=\frac{2}{\pi }\hbar
v_{F}K +  \qquad \qquad \qquad  & \nonumber & \\
 \left[ 1-\left\{ 2-\left( \frac{4}{\pi }\right) ^{2}\right\} \frac{%
E_{F}}{B_{2}}\right] \frac{\hbar ^{2}K^{2}}{2(2m)}+O(K^{3}), 
&\label{dk2} &
\end{eqnarray}
where a nonnegative CP {\it excitation energy }$\varepsilon _{K}$ has been
defined, and the Fermi velocity $v_{F}$ is given by $E_{F}/k_{F}=\hbar
v_{F}/2$. It is this excitation energy that must be inserted into the
Bose-Einstein (BE) distribution function to determine the critical
temperature in a picture of superconductivity (or of superfluidity in, e.g., liquid $^3$He) 
as a BE condensation (BEC) of CPs \cite{pla,Blatt}.

The leading term in (\ref{dk2}) is {\it linear} in the CMM, followed by a 
{\it quadratic} term; similar results hold in 3D. In the strong coupling limit ($B_{2}={\Delta _{0}}\gg E_{F}$) 
the quadratic term is exactly the CM kinetic energy of what was originally
a CP (becoming what is sometimes called a ``local pair''), namely, 
\begin{equation}
\lim\limits_{\Delta _{0}\gg E_{F}}\varepsilon {_{K}}=\frac{\hbar ^{2}K^{2}}{%
2(2m)},  \label{scl}
\end{equation}
the familiar nonrelativistic kinetic energy of the composite pair of mass $2m
$ and CMM $K$ in vacuum. It is {\it this} dispersion relation that has been
assumed in virtually all BEC studies in 3D of superconductivity (see, e.g., 
\cite{Blatt,Hauss} among others)---but which in 2D is well-known to give {\it zero}
transition BEC temperature $T_{c}$. However, recent BCS-Bose crossover picture
root-mean-square radii calculations compared with empirical coherence
lengths of several typical 2D-like cuprates \cite{1964} suggest these
superconductors to be well within the BCS or weak-coupling regime, implying
that the (nearly) linear relation is appropriate for them rather than the quadratic one. \ 

Fig. 1 displays exact numerical results obtained from (\ref{8}) for different couplings
of a dimensionless CP excitation energy $\varepsilon {_{K}/\Delta }_{0}$ as
function of $K/k_{F}$. For weak enough coupling (and/or high enough density) the {\it exact} dispersion relation is practically linear over almost the entire interval of $K/k_{F}$ up to breakup---despite
the divergence in (\ref{dk2}) of the quadratic and
possibly higher-order terms as $B_{2}\rightarrow 0$. For stronger coupling the quadratic dispersion
relation (\ref{scl}) begins to dominate, i.e., as $B_{2}/E_{F}\rightarrow
\infty $, meaning that either $\ B_{2}\rightarrow \infty $ {\it or} $%
E_{F}\rightarrow 0$, the latter implying that the Fermi sea vanishes and the
vacuum is recovered {\it as a limit}. \ 

In conclusion, the CP problem with nonzero CMM evolves with weak to
strong zero-range pairwise interaction to give a {\it linear} dispersion
relation in the CMM that gradually {\it crosses over} to a {\it quadratic }%
relation with increasing coupling. These results will play a critical role
in a model of superconductivity (or superfluidity) based on BE condensation of CPs,
particularly in 2D where only the linear relation can give rise to {\it nonzero }$T_{c}$%
's.

Partial support from UNAM-DGAPA-PAPIIT (M\'{e}xico) \# IN102198, CONACyT (M%
\'{e}xico) \# 27828 E, DGES (Spain) \# PB95-0492 and FAPESP\ (Brazil) is
gratefully acknowledged.

\vspace{-1.0cm}
\vspace{5cm}
Figure Caption

{1. Dimensionless exact CP excitation energy $\varepsilon _{K}/\Delta
_{0}$ {\it vs} $K/k_{F}$ in 2D (full curves) calculated from (\ref{8}) for
different $B_{2}/E_{F}$. Dot-dashed line is linear approximation
(virtually coincident with the exact curve for all $B_{2}/E_{F} \; 
\raisebox{-0.7ex}{${\stackrel{<}{{\small \sim}}}$} \; 0.1$) while dashed
curve is exact finite range result for a potential more general than
(\ref{1}) and whose double Fourier transform is $V_{pq} \equiv
-(v_0/L^2)[(1+p^2/p_0^2)(1+q^2/p_0^2)]^{-1/2}$ (see Nozi\`{e}res and
Schmitt-Rink \cite{Blatt}), which for $p_0 \to \infty$ becomes (\ref{1}).
Dots mark values of CMM wavenumber where the CP\ breaks up, i.e., where
$\Delta _{K}$ turns negative.} \label{fig:largenenough} 

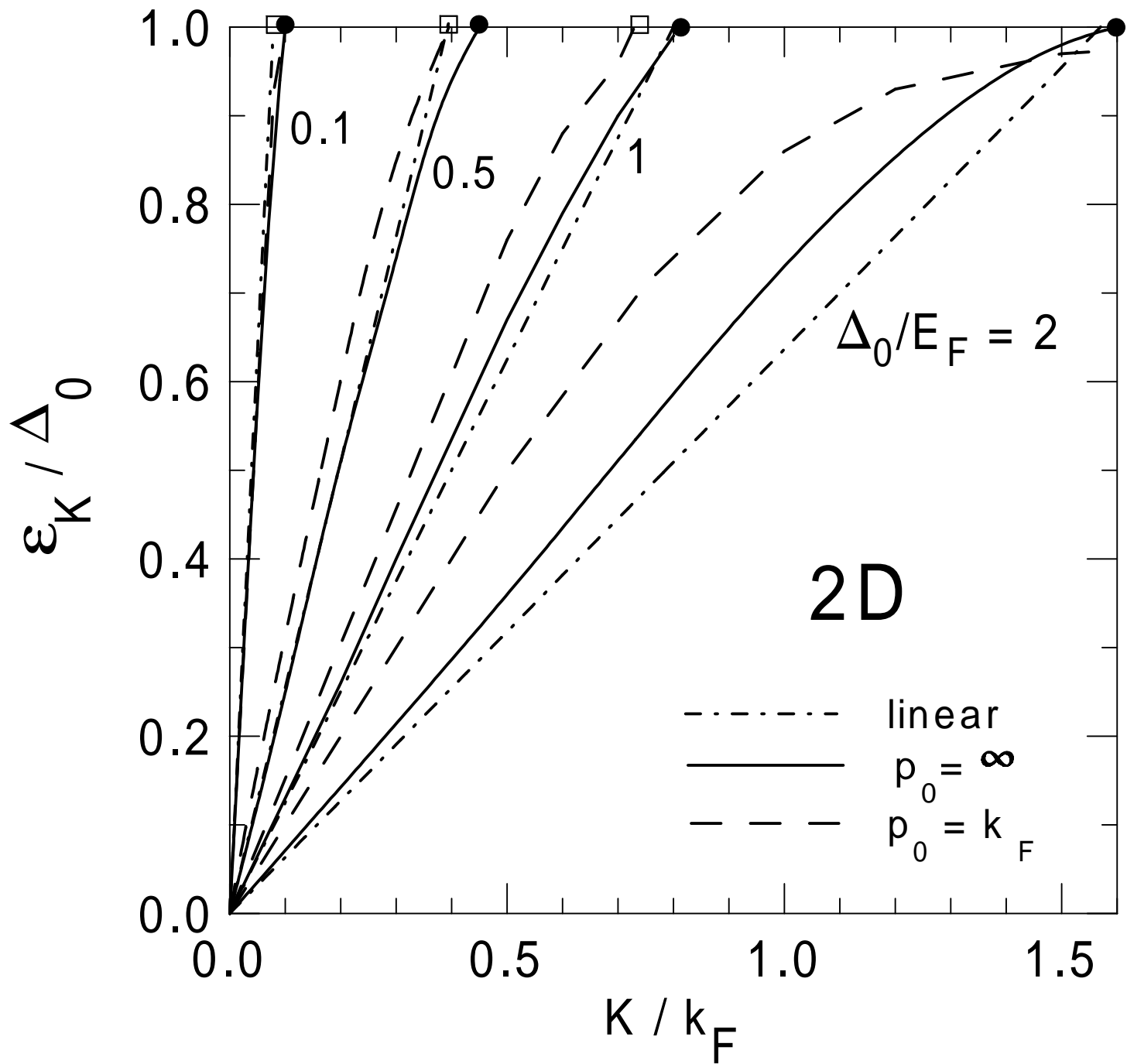
\end{document}